\documentclass[12pt]{iopart}
\usepackage{iopams}
\expandafter\let\csname equation*\endcsname\relax
\expandafter\let\csname endequation*\endcsname\relax

\usepackage{graphicx,amsmath,bm, dsfont,bbold,amssymb,amsfonts,latexsym,color,dcolumn,cite,epsfig}

\def\bra#1{\mathinner{\langle{#1}|}}
\def\ket#1{\mathinner{|{#1}\rangle}}

\def\text#1{\textrm{#1}}
\def\id{\mathds{1}}

\def\eq{\begin{equation}}
\def\eeq{\end{equation}}

\def\D{\mathcal{D}}

\begin{document}

\title{Comparing different approaches for generating random numbers device-independently using a photon pair source}%

\author{V.~Caprara Vivoli}
\address{Group of Applied Physics, University of Geneva, CH-1211 Geneva 4, Switzerland}
\author{P.~Sekatski}
\address{Institut for Theoretische Physik, Universitat of Innsbruck, Technikerstr. 25, A-6020 Innsbruck, Austria}
\author{J.-D.~Bancal}
\address{Centre for Quantum Technologies, National University of Singapore, 3 Science Drive 2, Singapore 117543}
\author{C.C.W.~Lim}
\address{Group of Applied Physics, University of Geneva, CH-1211 Geneva 4, Switzerland}
\author{A.~Martin}
\address{Group of Applied Physics, University of Geneva, CH-1211 Geneva 4, Switzerland}
\author{R.T.~Thew}
\address{Group of Applied Physics, University of Geneva, CH-1211 Geneva 4, Switzerland}
\author{H.~Zbinden}
\address{Group of Applied Physics, University of Geneva, CH-1211 Geneva 4, Switzerland}
\author{N.~Gisin}
\address{Group of Applied Physics, University of Geneva, CH-1211 Geneva 4, Switzerland}
\author{N.~Sangouard}
\address{Department of Physics, University of Basel, CH-4056 Basel, Switzerland}

\begin{abstract}
What is the most efficient way to generate random numbers device-independently using a photon pair source based on spontaneous parametric down conversion? We consider this question by comparing two implementations of a detection-loophole-free Bell test. In particular, we study in detail a scenario where a source is used to herald path-entangled states, i.e. entanglement between two spatial modes sharing a single photon and where non-locality is revealed using photon counting preceded by small displacement operations. We start by giving a theoretical description of such a measurement. We then show how to optimize the Bell-CHSH violation through a non-perturbative calculation, taking the main experimental imperfections into account. We finally bound the amount of randomness that can be extracted and compare it to the one obtained with the conventional scenario using photon pairs entangled e.g. in polarization and analyzed through photon counting. While the former requires higher overall detection efficiencies, it is far more efficient in terms of the entropy per experimental run and under reasonable assumptions, it provides higher random bit rates.
\end{abstract}
\date{\today}
\maketitle

\section{Introduction.}
In the last decades, the idea of using the randomness present in quantum phenomena to create random number strings has been pushed forward \cite{Jennewein00,Stefanov00,Dynes08}. Among the quantum techniques that are envisaged to expand a given random bit string, those based on a Bell test \cite{Pironio10, Colbeck11, Brunner14}, the so-called device-independent quantum random number generators (DI-QRNG), are very attractive because they are based on a few assumptions that are relatively easy to check in real time. The price to pay is to realize a Bell test without the detection loophole. The detection loophole has been addressed in several experiments including single ions \cite{Rowe01,Matsukevich08} and single atoms \cite{Hofmann12} and very recently, using photon pair sources \cite{Christensen13,Giustina13}. The latter has several advantages in practice in that it is much less restrictive in terms of wavelength and bandwidth than atoms. It further has the advantage of simple implementation since $\chi^{(2)}$ non linear crystals are well integrated devices, commercially available and operating at room temperature. The bottleneck of photonic experiments is the detector inefficiency, but given recent improvements \cite{Lita08,Miller11,Fukuda11, Verma14}, setups based on spontaneous parametric down conversion (SPDC) sources are attracting more and more attention, including for their commercial perspectives.\\

\begin{figure}[h]
\begin{center}
\includegraphics[width=260pt]{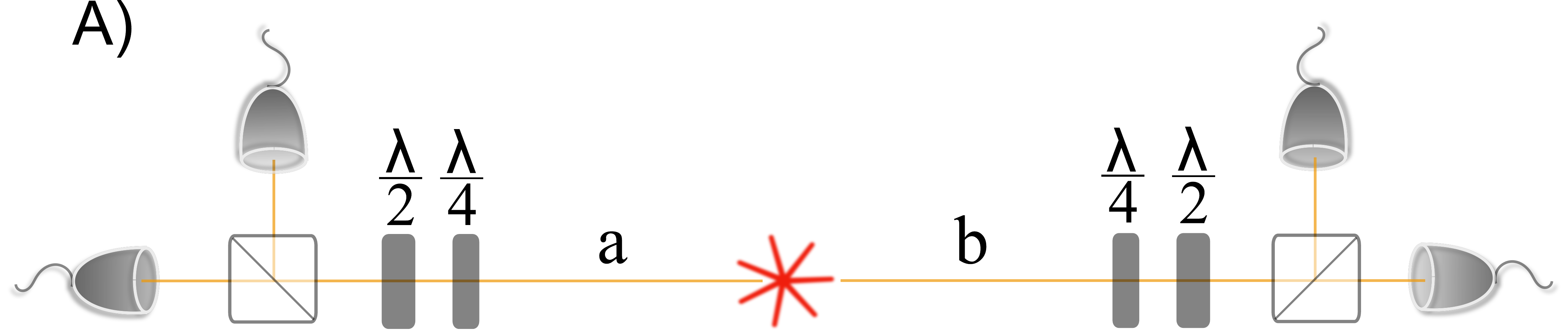}
\includegraphics[width=255pt]{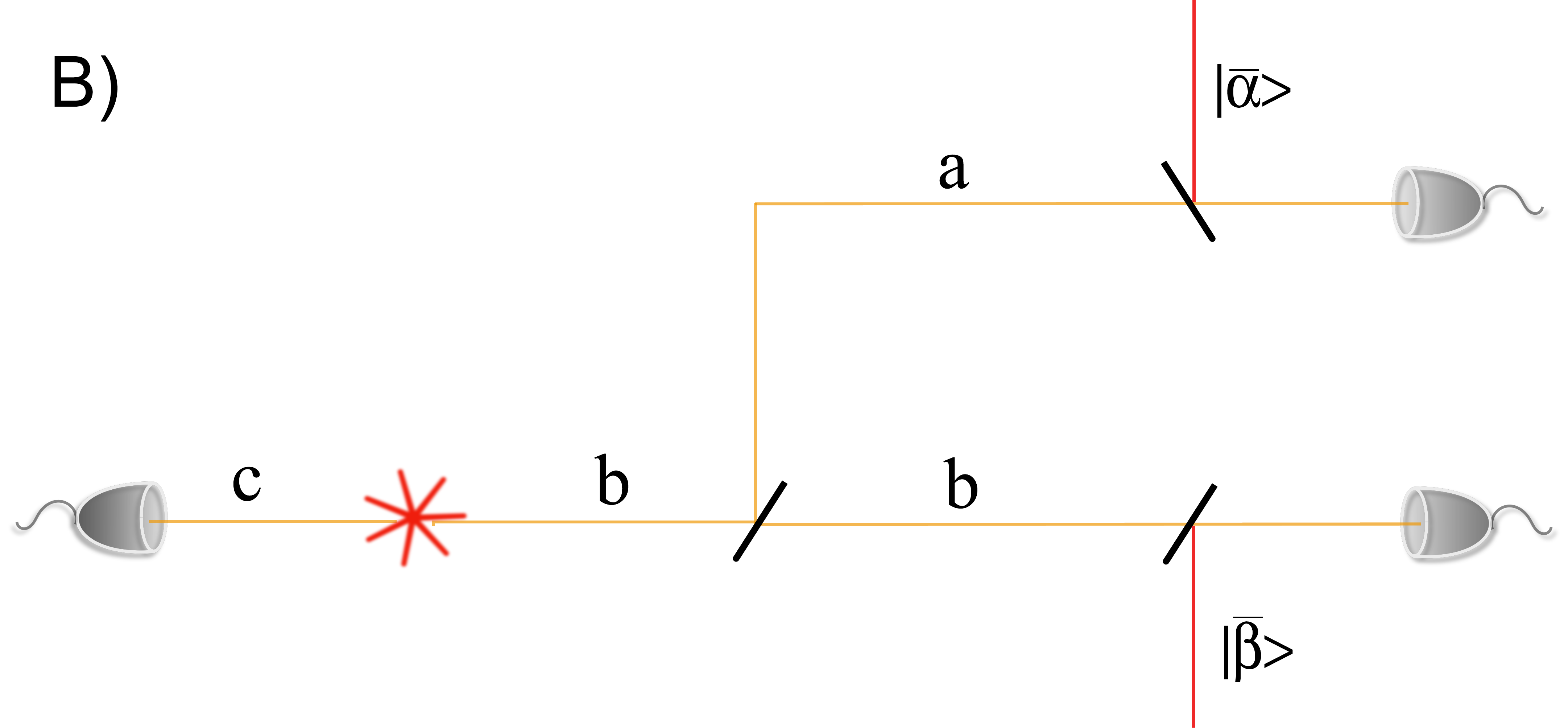}
\end{center}
\caption{Scheme of two possible implementations of a Bell test using a photon pair source. A) A source (star) based on SPDC is excited e.g. by a pulsed pump and produce photon pairs entangled e.g. in polarization. The photons are emitted in correlated spatial modes $a$ ($b$). Each of them might include several temporal/frequency/spatial modes $a_k$ --- $b_k$. The photons emitted in $a$ ($b$) are sent to Alice's (Bob's) location where they are projected along an arbitrary direction of the Bloch sphere using a set of wave-plates, a polarization beam splitter and two detectors. B) A source (star) based on SPDC produces photon pairs. We assume that in this scenario the emission is mono-mode. The detection of one photon thus heralds the creation of its twin in a pure state. The latter is sent through a beam splitter. This leads to an entangled state between the two paths a and b. The state of each path is displaced in the phase space using an unbalanced beamsplitter and a coherent state, before being detected though photon counting techniques. The detectors are assumed to be non-photon number resolving with non-unit efficiency.\label{Figure1}}
\end{figure}

The conventional setup, used e.g. in the experiments \cite{Giustina13,Christensen13}, is shown in Fig. \ref{Figure1}A). A SPDC source produces photon pairs entangled e.g. in polarization. The photons are then analyzed by a set of wave plates and non photon number resolving (NPNR) detectors\footnote{ Note that TES detectors are capable of number resolution. Nevertheless, this capability was not used in \cite{Giustina13,Christensen13}.}. Importantly, it has been realized recently \cite{Caprara14} that the maximal CHSH-Bell violation \cite{Clauser69} that can be reached in this scenario is intrinsically limited by the characteristics of the source, i.e. by the presence of vacuum and multiple photon pairs. As shown in Ref. \cite{Pironio10}, the observed CHSH violation can be used to quantify the amount of extractable randomness in the experimental data. That is, the min-entropy of the data is lower bounded by a function monotonically increasing in the observed CHSH violation. A reduction in the violation thus implies a reduction in the amount of extractable randomness. This raises the question of whether other scenarios involving similar resources could provide larger Bell violations and hence would be more suited for DI-QRNG.\\

An alternative scenario for Bell test with photons has been proposed by Banaszek and Wodkiewicz in 1998 \cite{Banaszek98} (see also related theoretical investigations \cite{Tan91,Chavez11,BohrBrask13,Torlai13,Seshadreesan13}) leading to a proof of principle experiment in 2004 \cite{Hessmo04}. The corresponding implementation using a SPDC source is shown in Fig. \ref{Figure1}B). A non linear crystal is pumped by a pulsed laser with an intensity carefully tuned to create a pair of photons with a small probability in modes b and c. A detection in c, even with an inefficient NPNR detector, heralds the creation of its twin photon in b. The latter is subsequently sent through a beam splitter, entangling the two output spatial modes a and b. Each of these modes is then analyzed through photon counting preceded by small displacements in phase space. Such a displacement is easily implemented in practice, using an unbalanced beamsplitter and a coherent state. In the subspace with at most one photon $\{\ket{0},\ket{1}\},$ this measurement corresponds to a noisy qubit measurement whose direction in the Bloch sphere depends on the size of the displacement, as detailed below. By choosing the appropriate settings and by taking the events "click" and "no-click" as binary outputs of a Bell test, a CHSH-Bell value of $\approx2.69$ can be obtained with a state of the form $\frac{1}{\sqrt{2}}(\ket{01}+\ket{10})$ \cite{Chavez11,BohrBrask13}. However, it was not previously clear what the maximum violation could be in a realistic scenario involving a SPDC source, non-unit efficiency and noisy detectors. Here we present such an analysis with the aim of establishing the best experimental setup for DI-QRNG. More precisely, we start by providing a detailed theoretical analysis of this measurement involving photon counting preceded by a small displacement operation. We then show how to calculate the Bell correlations in a non-perturbative way in the scenario presented in Fig. \ref{Figure1}B) that we call "spatial entanglement" in the rest of the paper. We then optimize the CHSH violation for a given detection efficiency $\eta$ over the squeezing parameter, the displacement amplitudes, and the splitting ratio of the beam splitter. Lastly we calculate the min entropy and the rate of random bits that can be extracted in this setup. We compare them to the conventional case where entangled pairs are detected by photon counting (see Fig. \ref{Figure1}A)). We show that while the scenario based on spatial entanglement requires higher overall detection efficiencies, it is preferable to the two photon case regarding the min entropy and, under reasonable assumptions, regarding the rate of random bits as well.


\section{Measurement analysis.}
In this section, we provide a detailed analysis of the measurement device used in the scenario based on spatial entanglement. We consider a NPNR detector of efficiency $\eta$ preceded by a displacement $\alpha = |\alpha| e^{i \delta}.$ The no click/click events are associated to two elements of a POVM $\{P_0, P_c\}$ which satisfy $P_0+P_c=\id.$ The no-click event of our NPNR detector is described by the operator $(1-\eta)^{a^\dag a}.$ Taking the displacement into account, one gets  $P_0 = \D^\dag(\alpha) (1-\eta)^{a^\dag a} \D(\alpha)$. To gain insight on this measurement, let us restrict $P_0$ to the Hilbert space spanned by $\ket{0}=\binom{1}{0}$ and $\ket{1}=\binom{0}{1}$ where it takes the following matrix form
\begin{align}
P_0=
\left(
\begin{array}{cc}
e^{-\eta|\alpha|^2} & -\eta\, \alpha^* e^{-\eta|\alpha|^2}\\
-\eta \, \alpha e^{-\eta|\alpha|^2} & (1-\eta +\eta^2 |\alpha|^2) e^{-\eta|\alpha|^2}
\end{array}\right)
\end{align}
Let us recall that $P_c=\id - P_0$. For non-unit efficiency $\eta < 1,$ the POVM $\{ P_0, P_c\}$ is not extremal \cite{Ariano05}
\eq
\{ P_0, P_c\} = \mu \{ \Pi_{\vec n}, \Pi_{-\vec n}\} + (1-\mu)\{ r_0 \id, r_c \id \}.
\eeq
This means that this measurement corresponds to a projective measurement in the direction
$$
\vec n \propto
\left(\begin{array}{c}
- e^{-\eta |\alpha|^2} |\alpha| \eta \cos(\delta) \\
e^{-\eta |\alpha|^2} |\alpha| \eta \sin(\delta)\\
 \frac{1}{2} e^{-\eta |\alpha|^2}  \eta (1 -|\alpha|^2 \eta )
\end{array}\right)\\
$$
on the Bloch sphere with probability
$$
\mu = \sqrt{\eta ^2 e^{-2 |\alpha| ^2 \eta } \left(|\alpha| ^2 \left(  |\alpha| ^2 \eta^2 -2\eta+4\right)+1\right)}.
$$
\begin{figure}[h!]
\begin{center}
\includegraphics[width=6cm]{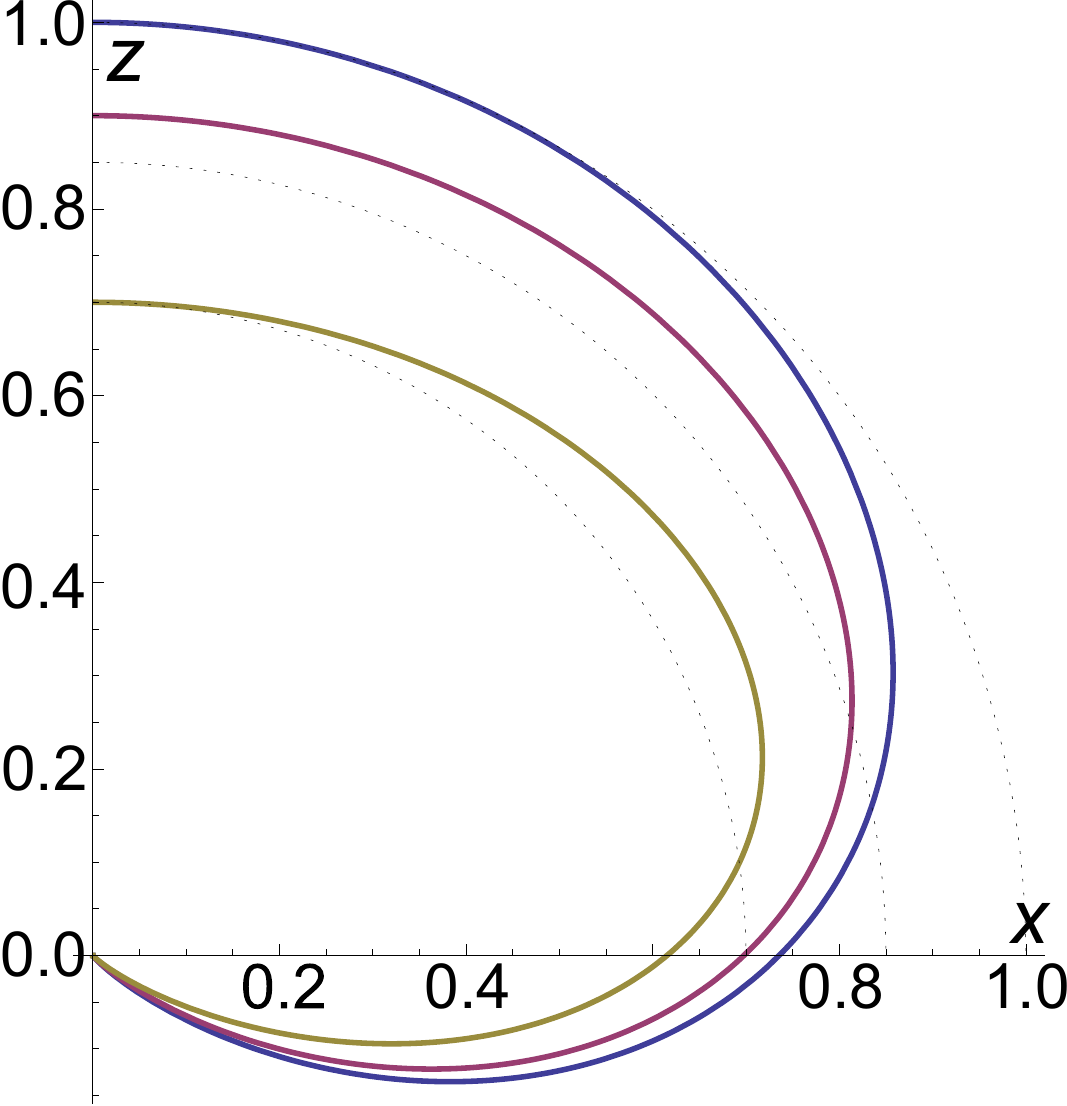}
\caption{Focusing on the projective part  $\mu\,  \{\Pi_{\vec n},\Pi_{-\vec n}\}$ of the studied POVM, we here represent the length $\mu$ and the direction of the corresponding vector in the Bloch sphere. As we consider real $\alpha,$ this vector lies in the x-z plane. For unit detection efficiency ($\eta=1,$ outermost curve) and $\alpha=0,$ this vector has a unit length and is directed to the z direction. When $\alpha$ increases ($\alpha$ spans the interval  $[0,4]$), the vector starts to rotate (the polar angle gives the azimuthal angle of $\vec n$ on the Bloch sphere) and its length decreases (the radius decreases). In the limit of large $\alpha,$ the vector length tends to zero. The two inner curves corresponds to non-unit efficiency ($\eta=90$\% and $\eta=70$\%, respectively). \label{fig_parametric}}
\end{center}
\end{figure}
With the remaining probability $(1-\mu),$ the output of the measurement is given randomly (regardless of the input state) accordingly to the distribution $\{ r_0, r_c\}$ where
\eq
r_0 =\frac{1}{2} \times\frac{\frac{\eta  \left(|\alpha| ^2 \eta -1\right)+2}{\sqrt{\eta ^2 \left(|\alpha| ^2 \left(\eta  \left(|\alpha| ^2 \eta -2\right)+4\right)+1\right)}}-1}{\frac{e^{|\alpha| ^2 \eta }}{\sqrt{\eta ^2 \left(|\alpha| ^2 \left(\eta  \left(|\alpha| ^2 \eta -2\right)+4\right)+1\right)}}-1}
\eeq
and $r_c=1-r_0$.
As an example, consider the case without displacement $\alpha=0$. The previous POVM reduces to
\eq
\{ P_0, P_c\} = \eta \{ \Pi_{0}, \Pi_{1}\} + (1-\eta)\{ \id, 0 \}
\eeq
i.e. it corresponds to a projective measurement in the direction z with the probability $\eta$ and with the remaining probability $(1-\eta),$ a no-click event occurs regardless of the input state. \\

Note that the phase term of the displacement $e^{i\delta}$ affects the polar angle of $\vec n$ only. For simplicity, we consider the case $\alpha=|\alpha|$ where the direction of the measurement lays in the x--z plane of the Bloch sphere. We further focus on the projective part of the POVM $\mu\,  \{\Pi_{\vec n},\Pi_{-\vec n}\}$ and we look at the direction and length of the corresponding vector $\mu \vec n$ on the Bloch sphere. The result is shown in Fig. \ref{fig_parametric}. For $\eta=1$ and $\alpha=0,$ this vector is directed in the z direction and has a unit length. The measurement device thus performs a projection along z. When $\alpha$ increases, the vector starts to rotate toward x while its length reduces. For non-unit efficiencies, the vector is shorter and it also rotates toward x when $\alpha$ increases. Surprisingly, we remark that the vector length increases with $\alpha$ (before it drops to zero), i.e. the ''effective detection efficiency" of the measurement setup $\mu$ gets larger than the intrinsic efficiency of the detector itself $\eta$.


\section{Exact Derivation of Bell-CHSH correlators.}
The purpose of this section is to derive the exact expression of the CHSH-Bell correlators in the case of spatial entanglement (see Fig. 1B)). We first focus on the density matrix $\rho_h$ of b resulting from a detection in c. The state  created by the SPDC source is given by $\ket{\psi}=\sqrt{1-T^2_g}\sum_n\frac{T^n_g}{n!}b^{\dagger n}c^{\dagger n}\ket{00}$, where $T_g=\tanh(g)$, $g$ being the squeezing parameter. 
To obtain $\rho_h,$ we have to calculate $tr_c\left(\ket{\psi}\bra{\psi}\left(\id-(1-\eta_h)^{c^{\dagger}c}\right)\right)$. $\eta_h$ stands for the efficiency of the heralding detector and $tr_c$ is the trace on $c$. This can be expressed as the difference of two terms. The first one is simply the trace over $\ket{\psi}$ while the second one can be written as $tr_c\left(R_h^{c^{\dagger}c}\ket{\psi}\bra{\psi}R_h^{c^{\dagger}c}\right)$, with $R_h=\sqrt{1-\eta_h}$. Using the formula $R_h^{c^{\dagger}c}e^{T_g a^{\dagger}c^{\dagger}}=e^{R_h T_g a^{\dagger}c^{\dagger}}R_h^{c^{\dagger}c}$ \cite{Sekatski10},  and re-normalizing the obtained state, the resulting density matrix $\rho_h$ can be written as
\begin{equation}
\rho_h=\frac{1-R_h^2T^2_{g}}{T^2_{g}\left(1-R_h^2\right)}\left[\rho_{\text{th}}\left(\bar{n}=\frac{T_g^2}{1-T_g^2}\right)-\frac{1-T_g^2}{1-R_h^2T_g^2}\rho_{\text{th}}\left(\bar{n}=\frac{R_h^2T_g^2}{1-R_h^2T_g^2}\right)\right],\label{EquThermalState}
\end{equation}
i.e. a difference between two thermal states $\rho_{\text{th}}(\bar{n})=\frac{1}{1+\bar{n}}\sum_{k}\left(\frac{\bar{n}}{1+\bar{n}}\right)^k\ket{k}\bra{k}$ where $\bar{n}$ is the mean photon number. Let us first calculate the correlators that would be obtained from a thermal state. We recall that a thermal state is classical with respect to the P representation. Therefore, it can be written as a mixture of coherent states $\ket{\gamma}.$ Concretely, $\rho_{\text{th}}\left(\bar{n}\right)=\int d^2 \gamma P^{\bar{n}}(\gamma)\ket{\gamma}\bra{\gamma}$
with $P^{\bar{n}}(\gamma)=\frac{1}{\pi \bar{n}}e^{-\frac{|\gamma|^2}{\bar{n}}}$. The correlators associated to a thermal state can thus be obtained by looking at the behaviour of a coherent state. A beam splitter splits a coherent state into two coherent states, i.e. $\ket{\gamma}\rightarrow\ket{\sqrt{R}\gamma}_a\ket{\sqrt{T}\gamma}_b$, where $T$ and $R$ are, respectively, the transmittivity and the reflectivity. A displacement $D(\alpha)$ on a coherent state $\ket{\gamma}$ gives another coherent state with mean photon number $|\gamma+\alpha|^2$, i.e. $D(\alpha)\ket{\gamma}=\ket{\gamma+\alpha}$.
From
\begin{equation}
\left(1-\eta\right)^{\frac{a^{\dagger}a}{2}}\ket{\bar{\gamma}}=e^{-\frac{\eta|\bar{\gamma}|^2}{2}}\ket{\sqrt{1-\eta}\bar{\gamma}},
\end{equation}
we easily obtain the probability to get no click in both sides from a thermal state $\rho_{\text{th}}(\bar{n})$ knowing the amplitudes of the local displacements $\alpha$ and $\beta$
\begin{equation}
p^{\text{nc,nc}}_{\alpha,\beta}=\frac{e^{-\eta\left(|\alpha|^2+|\beta|^2\right)+\frac{\bar{n}\eta^2}{1+\bar{n}\eta}|\sqrt{R}\alpha+\sqrt{T}\beta|^2}}{1+\eta \bar{n}}.
\end{equation}
Attributing the value $+1$ ($-1$) to a "no-click" event ("click" event), we then obtain an explicit expression for the correlator $E^{\text{th}}_{\alpha, \beta}=p^{\text{nc,nc}}_{\alpha,\beta}+p^{\text{c,c}}_{\alpha,\beta}-p^{\text{nc,c}}_{\alpha,\beta}-p^{\text{c,nc}}_{\alpha,\beta}$ associated to a thermal state $\rho_{\text{th}}(\bar{n})$
\begin{equation}
\nonumber
\begin{split}
E^{\text{th}}_{\alpha, \beta}
=&1+4\frac{e^{-\eta\left(|\alpha|^2+|\beta|^2\right)+\frac{\bar{n}\eta^2}{1+\bar{n}\eta}|\sqrt{R}\alpha+\sqrt{T}\beta|^2}}{1+\eta \bar{n}}-2\frac{e^{-\frac{\eta|\alpha|^2}{1+\eta \bar{n} R}}}{1+\eta\bar{n}R}-2\frac{e^{-\frac{\eta|\beta|^2}{1+\eta \bar{n} T}}}{1+\eta\bar{n}T}.
\end{split}\end{equation}
From this last expression, we deduce the correlator $E_{\alpha_i \beta_j}$ for the state (\ref{EquThermalState})
\begin{equation}
E_{\alpha_i \beta_j}=\frac{1-R_h^2T^2_{g}}{T^2_{g}\left(1-R_h^2\right)}\left[E^{\text{th}}_{\alpha_i, \beta_j}\left(\bar{n}=\frac{T_g^2}{1-T_g^2}\right)-\frac{1-T_g^2}{1-R_h^2T_g^2}E^{\text{th}}_{\alpha_i, \beta_j}\left(\bar{n}=\frac{R_h^2T_g^2}{1-R_h^2T_g^2}\right)\right].\label{EquThermalState2}
\end{equation}
This explicit expression of $E_{\alpha_i \beta_j}$ allows one to optimize the CHSH-Bell value, i.e. the value of $S=|E_{\alpha_1 \beta_1}+E_{\alpha_1 \beta_2}+E_{\alpha_2 \beta_1}-E_{\alpha_2 \beta_2}|$, for given efficiencies ($\eta$, $\eta_h$) over the tunable parameters of the system, i.e. the squeezing parameter $g$, the amplitude of the local displacements $\alpha_i$ and $\beta_j$ (measurement settings), and the transmittivity $T$ of the beam splitter. Note that the CH \cite{Clauser74} and CHSH inequalities are equivalent for all probability distributions satisfying the no-signaling condition, i.e. for all quantum correlations \cite{Brunner14}. Namely, they are related by the affine relation $4\, CH = S-2$.


\section{Optimization of the CHSH value.}
\begin{figure}[h]
\begin{center}
\includegraphics[width=230pt]{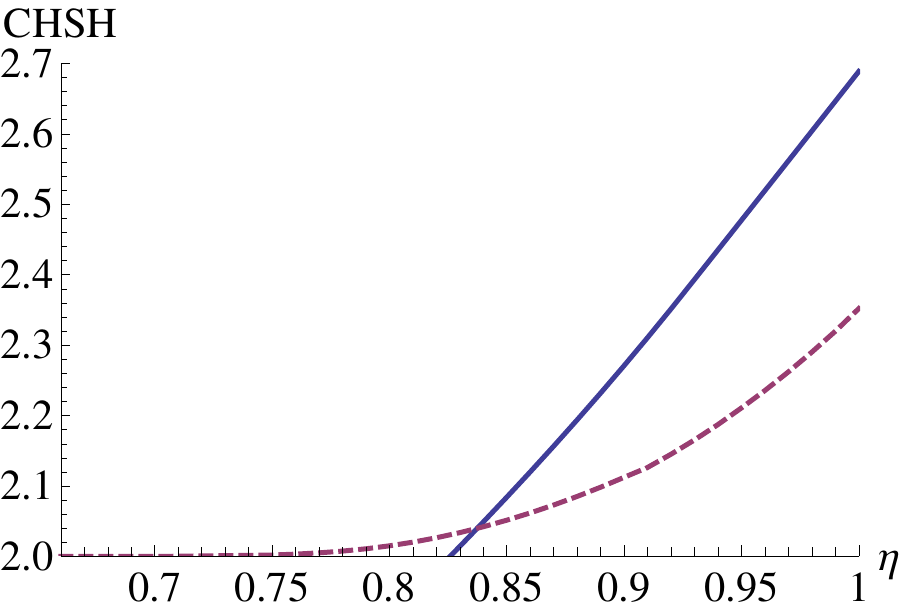}
\end{center}
\caption{Optimal CHSH value as a function of the efficiency $\eta$. The full (dashed) curve is obtained in the case of spatial entanglement, see Fig. \ref{Figure1}B) (polarization-entanglement, see Fig. \ref{Figure1}A)) (see the text for detail).\label{Figure2}}
\end{figure}
In this section, we present the result of the optimization of the CHSH-Bell value in the case of spatial entanglement (Fig. \ref{Figure1}B)). Fig. \ref{Figure2} shows S as a function of the efficiency $\eta$ and compares it to the case of polarization entanglement for which the optimization of the CHSH-Bell value has been reported in Ref. \cite{Caprara14}. We emphasize that $\eta$ is the overall detection efficiency including the transmission efficiency from the source to the detector. We assume that the efficiencies for modes a and b are the same. They are equal to the heralding efficiency $\eta_h=\eta$. Three results deserve to be highlighted.
\begin{enumerate}
\item[i.] In the ideal case where $\eta=1,$  the maximal violation is around $2.69$. This value is obtained in the limit $g\rightarrow 0$, i.e. when the production of multiple photon pairs is negligible. Since the heralding signal eliminates the vacuum component, we end up with a single photon Fock state in b to a very good approximation. We thus retrieve the maximal violation that can be obtained in the scenario presented in Fig.  \ref{Figure1}B) with a single photon \cite{Chavez11,BohrBrask13}. Note that in practice, the value of $g$ is limited by the probability $p_{dc}$ of having a dark count in the heralding detector which is negligible if $p_{dc} \ll \eta T_g^2$ only. More concretely, if one assumes that the probability of having a dark count is $p_{dc}\approx 10^{-5}$ for example, we found the optimal violation $S \sim 2.67$ which is obtained for $g \sim 0.07$ and still $\eta=1.$

\item[ii.] We observe that the optimal state is always obtained from a 50-50 beam splitter ($R=T=\frac{1}{2}$) in the limit $g \rightarrow 0$, i.e. it is a two-qubit maximally entangled state. This is unexpected as in the case of a two-qubit state entangled in polarization, lower efficiencies can be tolerated from non-maximally entangled states \cite{Eberhard93}. 

\item[iii.] The minimal required detection efficiency is $\eta_{\text{min}}=0.826$. This is counterintuitive, at least at first sight, since there is a local model reproducing the correlation of the singlet state as soon as the detection efficiency is lower or equal to $\frac{2}{\sqrt{2}+1}\approx 0.828$ \cite{Garg87,Larsson98,Massar02}. Nevertheless, this model assumes that the probability of having a conclusive event is $\eta$ whereas the probability for having a non conclusive event is $1-\eta.$ This does not hold in the case of spatial entanglement. Let us also recall that in the scenario of spatial entanglement, the effective efficiency of the overall measurement device can be higher than the detection efficiency of the NPNR detector.

\end{enumerate}
Note that the CHSH-Bell values given in Fig. \ref{Figure2} are optimized over the local strategies that are used to assign binary results $\pm1$ to physical events (click and no-click). We found that they are all equivalent, i.e. they all lead to the same value of S. The sum $E_{\alpha_1 \beta_1}+E_{\alpha_1 \beta_2}+E_{\alpha_2 \beta_1}-E_{\alpha_2 \beta_2}$ simply needs to be minimized or maximized depending on the strategy.


\section{Rate of random bit generation.}
\begin{figure}[h]
\begin{center}
\includegraphics[width=230pt]{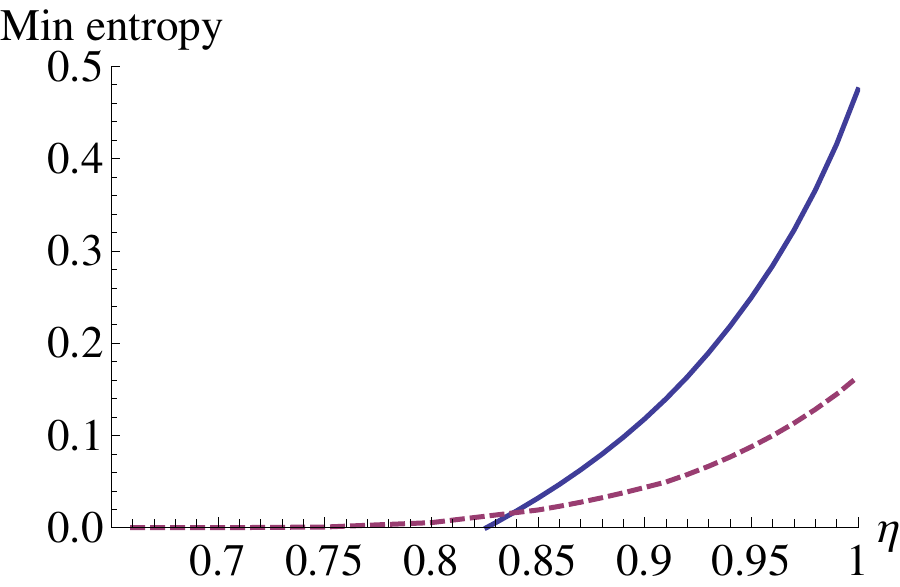}
\end{center}
\caption{Min entropy per experimental run as a function of the efficiency $\eta$. The full (dashed) curve corresponds to the case of spatial entanglement (Fig. \ref{Figure1}b)) (polarization-entanglement, see Fig. \ref{Figure1}a)).\label{Figure3}}
\end{figure}

In this section, we estimate the amount of randomness created in both setups that are presented in Fig. \ref{Figure1}. We present two quantities, (i) the randomness per run, i.e. the min entropy $H_{\text{min}}(S)$, and (ii) the rate of randomness generation. Let us first focus on the min entropy $H_{\text{min}}(S)$. As mentioned earlier in the introduction, the min-entropy rate (amount of randomness per bit) can be lower bounded in terms of the observed CHSH violation S \cite{Pironio10}. The lower bound is given by
\begin{equation}
H_{\text{min}}(S)=1-\log_2\left(1+\sqrt{2-\frac{S^2}{4}}\right).
\end{equation}
$H_{\text{min}}(S)$ is equal to $0$ when $S$ is $2$ and it reaches its maximum value $1$ when $S$ is maximal, i.e. $S=2\sqrt{2}$ \cite{Bancal13}.
Since the min entropy is a monotonic function of $S$, large $S$ favors large min entropy. The optimal value of $H_{\text{min}}(S)$ computed from \cite{Pironio10} for the two different implementations of Fig. \ref{Figure1} is shown in Fig. \ref{Figure3}. Since a larger violation can be obtained in the scenario involving spatial entanglement, the scheme of Fig. \ref{Figure1}B) provides higher min entropy than the scheme of Fig. \ref{Figure1}A) for large enough efficiencies. On the other hand, the scenario involving the spatial-entanglement requires efficiencies higher than $0.826$ while the scenario with polarization-entangled states allows one to get small but non-zero min entropy for efficiencies in between $\approx0.67$ and $0.826$.\\

\begin{figure}[ht]
\begin{center}
\includegraphics[width=230pt]{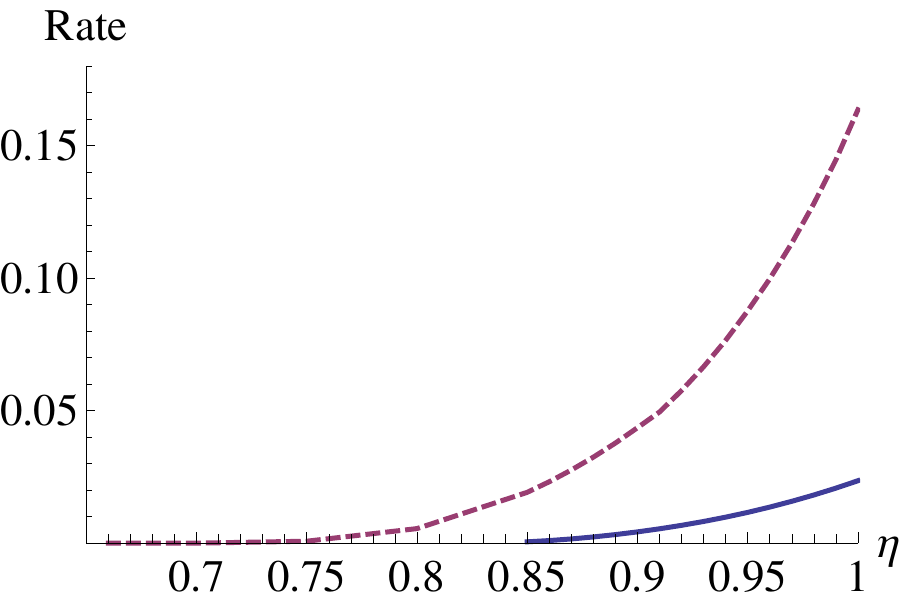}
\end{center}
\caption{Rates of random bits (in unit of the pump-laser rate) as a function of the efficiency $\eta$. The full (dashed) curve corresponds to the case of spatial-entanglement (polarization-entanglement).\label{Figure4}}
\end{figure}

Let us now focus on the rate of randomness generation. It is given by
\begin{equation}
R(S)=r H_{\text{min}}(S)
\end{equation}
where $r$ is the rate at which the states are analyzed. Consider first the case where the repetition rate is set by the pump laser. For the conventional setup (Fig. \ref{Figure1}A)) $R(S)=r_{\text{pump}} H_{\text{min}}(S)$ whereas in the case of spatial entanglement, the rate at which the states are analyzed is intrinsically limited by the heralding rate, i.e. $R(S)=r_{\text{pump}}\frac{\eta_hT^2_g}{1-(1-\eta_h)T^2_g}H_{\text{min}}(S)$. Assuming $\eta_h=\eta$, we have optimized $R(S)$ over the squeezing parameter $g$, the values of $\alpha_i$ and $\beta_j$, and the transmittivity $T$. The result is shown in Fig. \ref{Figure4} and is compared to the conventional scenario (see Fig. \ref{Figure1}A)). One sees that the high violations that are obtained in the scenario involving the spatial entanglement do not compensate the reduction of the repetition rate. \\

Consider now the situation where the rate is not limited by the pump laser but by the speed at which the measurement settings are chosen, as in Ref.\cite{Christensen13}, or by the deadtime of the detectors so that the heralding rate ($r_d$) in the scenario given in Fig. \ref{Figure1}A) is the same that the detection rate of the scenario of Fig. \ref{Figure1}B). In this case, the rate of random bits is given by $R(S)=r_d H(S)$ and can thus be deduced from Fig. \ref{Figure3}. It is clear that the rate of randomness in the scenario involving spatial entanglement is substantially higher than the conventional one  (Fig. \ref{Figure1}A)) for efficiencies larger than $0.84$ as its Bell violation is higher. Furthermore, in practice, randomness extraction is normally carried out on a fixed input bit string and the size of the output string is approximately given by the min-entropy of the input bit string. Seen from this point of view, it is clear that our spatial entanglement setup allows a larger number of extractable secret bits for a fixed input bit string. \\


\section{Conclusion \& Discussions.}
Motivated by very recent experiments reporting on the first-detection-loophole-free Bell tests with photon pairs, we have studied two different scenarios, both of them based on SPDC sources and photon counting techniques, for the generation of random bits. In particular, we have shown how to calculate the correlators in the scenario involving spatial entanglement (represented in Fig. \ref{Figure1}B)) in a non-perturbative way. This allowed us to optimize the CHSH-Bell value, that we have compared to the one obtained in the more conventional scenario of Fig. \ref{Figure1}A). While the detection technique of the scenario given in Fig. \ref{Figure1}B) involves small displacement operations, i.e. requires a noise free local oscillator indistinguishable from the photons to be detected, and overall detection efficiencies larger than in the conventional scenario, the scenario involving spatial entanglement has several interesting features:
\begin{itemize}
\item[i.] First, only one mode needs to be detected efficiently. Therefore one can use filtering techniques on the heralding mode to prepare it in a mode having high coupling and detection efficiency \cite{Pomarico12, Guerreiro13}.
\item[ii.] For efficiencies higher than $84 \%$, the scenario based on spatial entanglement leads to substantial improvements over the conventional setup in terms of min entropy.
\item[iii.] Assuming that the number of experimental runs is large enough so that the Bell violation is accurately estimated in both setups, we have shown that in the realistic regime where the repetition rate is limited e.g. by the detector dead time in both scenarios, the higher CHSH-Bell violation of the scenario with spatial entanglement leads to higher bit rates than the one of the conventional scenario.
\end{itemize}
We believe that these advantages could provide motivations for several experimental research groups to realize detection-loophole free Bell tests following the idea that K. Banaszek and K. Wodkiewicz \cite{Banaszek98} have initiated more than 15 years ago.


\section{Acknowledgements.}
We thank V. Scarani, T.Barnea, and G.P\"utz for discussions and comments. This work was supported by the Swiss NCCR QSIT, the Swiss National Science Foundation SNSF (grant PP00P2$\_$150579 and "Early PostDoc.Mobility"), the European Commission (IP SIQS, Chist-era DIQIP), the Singapore Ministry of Education (partly through the Academic Research Fund Tier 3 MOE2012-T3-1-009) and the Singapore National Research Foundation.\\

\end{document}